\documentclass[twocolumn,epsfig,aps,prb]{revtex4}

\begin{document}

\title{Level of holographic noise in interferometry.}

\author{Igor I. Smolyaninov}
\affiliation{Department of Electrical and Computer Engineering, University of Maryland, College Park, MD 20742, USA}

\date{\today}

\begin{abstract}
The level of holographic noise expected to be observed in interferometric gravitational wave detectors such as GEO600 is re-examined. It is demonstrated that earlier estimates are based on assumed linear diffractive behavior of Planck radiation. Since nonlinear effects, such as self-focusing, are expected to appear at much lower energies, the expected level of holographic noise must be reduced by many orders of magnitude.      
\end{abstract}


\maketitle

Quantum geometrical effects are expected to introduce uncertainty in precise measurements of length and time intervals. These effects may manifest themselves as a new source of noise in interferometers and other experimental setups \cite{1,2}. Recent work by Hogan \cite{3,4} indicates that these effects may be surprisingly close to being observable at currently achieved sensitivity of interferometric gravitational wave detectors, such as GEO600 \cite{5}. Moreover, \lq\lq holographic noise \rq\rq predicted by Hogan was suggested as a source of the unexplained noise currently observed at GEO600 \cite{4}. Such a surprising claim requires considerable theoretical and experimental scrutiny. Since the predicted level of holographic noise seems to be a direct consequence of the holographic hypothesis without any free parameters \cite{3}, the critical examination of the holographic noise becomes even more urgent.

The derivation of the holographic noise presented in ref.\cite{3} appears to be the most straightforward to analyze. It is based on the linear diffraction and Rayleigh criterion. A metric where the positions of events and paths are defined using only waves longer than a cutoff $\lambda_p$ is considered. It is demonstrated that within the scope of a linear theory the paths connecting events are then subject to large indeterminacy because of the limitation of defining the end points of a ray, or a path due to diffraction. If one end point of the particle path is limited within an aperture of size $D$, the other end point is uncertain within a diffraction spot of size $\sim \lambda_pL/D$ upon propagation over length $L$. If we wish to minimize the range of possible orientations consistent with these end points, we need to chose $D \sim \lambda_pL/D$. This minimization corresponds to the uncertainty of the end point positions within $\Delta x\sim (\lambda_pL)^{1/2}$. Hogan asserts that the orientation of a ray of wavelength $\lambda $ over a length $L$ can at best be defined with a precision of $\Delta \theta \sim (\lambda/L)^{1/2}$ and calls it \lq\lq an unavoidable classical transverse indeterminacy of rays that are defined by waves \rq\rq. He proceeds to conjecture that the transverse
indeterminacy of Planck-wavelength quantum paths corresponds to quantum indeterminacy of the metric itself. The commutation relation 
[$x(z_1),x(z_2)$]=$-il_p(z_2-z_1)$ is derived based on the linear diffraction of Planck radiation over the length $L=(z_2-z_1)$ (see eqs. (1-3) from ref.\cite{3}). 

The weak point of the holographic noise derivation is the assumption of \lq\lq classical unavoidability \rq\rq of ray diffraction. Modern optics demonstrates numerous examples of experimental situations in which the Rayleigh criterion and the diffraction limit are violated. This happens both in linear \cite{6,7}  and non-linear optics \cite{8,9,10}. These observations indicate that both end points of a ray may be localized in sub-wavelength regions of space upon propagation over arbitrary large ($L>>\lambda $) distances. The simplest example is self-focusing of the optical beams \cite{8,9}. Self-focusing is observed when radiation propagates through nonlinear media. Several physical mechanisms may produce variations in the refractive index of the material, which result in self-focusing. The most well known cases include self-focusing due to Kerr effect and self-focusing in plasmas. Self focusing is described by introduction of nonlinear refractive index as $n = n_0 + n_2I$, where $n_0$ and $n_2$ are the linear and non-linear components of the refractive index, and $I$ is the radiation intensity. In most materials $n_2$ is positive (the opposite case leads to self-defocusing). Therefore, the refractive index becomes larger at the centre of a beam, creating a focusing refractive index profile. Regardless of their initial shape, self-focusing beams are known to evolve into the same Townes profile \cite{11}.

Vacuum is known to behave as a nonlinear optical medium at energies of the order of $m_ec^2$, much below the Planck scale \cite{12}. While exact behavior of the nonlinear refractive index of vacuum as a function of energy is not known, the positive sign of $n_2$ is required by causality. Hence, self-focusing of vacuum at $\lambda \sim l_p$ should be expected. Therefore, the ray orientation uncertainty of $\Delta \theta \sim (\lambda_p/L)^{1/2}$ and the corresponding commutation relation [$x(z_1),x(z_2)$]=$-il_p(z_2-z_1)$ do not seem to be well justified. Instead, we must conclude that the ray orientation uncertainty must be of the order of $\Delta \theta \sim (\lambda_p/L)$, which is 19 orders of magnitude smaller. 

In conclusion, Hogan's method only works if self-focusing effects are negligible in the effective theory of holographic geometry.

\end{document}